\documentclass[letterpaper,twocolumn,10pt]{article}
\usepackage{usenix2019_v3}
\pdfoutput=1 

\usepackage{tikz}
\usepackage{amsmath}
\usepackage{amssymb} 
\usepackage{booktabs}
\usepackage{graphicx} 
\usepackage{color,soul}
\graphicspath{ {./images/} }
\begin{document} 

\date{}

\title{\Large \bf Cross-Router Covert Channels}

\author{{\rm Adar Ovadya, Rom Ogen, Yakov Mallah, Niv Gilboa and Yossi Oren }\\
Faculty of Engineering Sciences, Ben-Gurion University of the Negev\\
\texttt{\{adarov@post.| romog@post.| mallah@post.| gilboan@| yos@\} bgu.ac.il}
} 

\maketitle

\begin{abstract}
Many organizations protect secure networked devices from non-secure networked
devices by assigning each class of devices to a different logical network. These
two logical networks, commonly called the host network and the guest network,
use the same router hardware, which is designed to isolate the two networks in
software.

In this work we show that logical network isolation based on host and guest
networks can be overcome by the use of cross-router covert channels. Using
specially-crafted network traffic, these channels make it possible to leak data
between the host network and the guest network, and vice versa, through the use
of the router as a shared medium. We performed a survey of routers representing
multiple vendors and price points, and discovered that all of the routers we
surveyed are vulnerable to at least one class of covert channel. Our attack can
succeed even if the attacker has very limited permissions on the infected
device, and even an iframe hosting malicious JavaScript code can be used for
this purpose. We provide several metrics for the effectiveness of such channels,
based on their pervasiveness, rate and covertness, and discuss possible ways of
identifying and preventing these leakages. 

\end{abstract}

\section{Introduction}
Network separation and network isolation are important components of the
security policy of many organizations. The goal of these policies is to prevent
network intrusions and information leakage by separating sensitive network
segments from other segments of the organizational network, and indeed from the
general Internet. The traffic sent over the sensitive network segments may
include mission-critical business documents, control data for industrial
systems, or private health records. Less sensitive data may include multimedia
streams, environmental sensor readings or data related to the operation of home
automation devices. 

The different levels of security also extend to the networked devices
themselves. While some devices are protected from security risks by their owners
and manufacturers, either by careful administration or by the use of automatic
updates, other networked devices, such as Internet of Things (IoT)
nodes~\cite{DBLP:conf/soca/ZhangCWHCS14} or medical
devices~\cite{DBLP:journals/cacm/SametingerRLO15}, are difficult or impossible
to patch, and are considered to be at a higher risk of malware infection. It is
especially important to isolate these less-secure networked devices from other
devices on the network.

A common approach for achieving network isolation is to {\em logically} separate
one physical network into multiple logical networks. Many routers provide 
this functionality by splitting the network into a \emph{host network} and a
\emph{guest network}. The router discards any traffic traveling between one
network and the other, enforcing separation as long as nodes on the two networks
do not connect to a common node on the Internet.

Logical isolation is not only common in practice, but it is actively recommended as a
security measure. For example, the U.S. National Institute of Standards and
Technology (NIST) \cite{stouffer2011guide} recommends isolating Industrial
Control System components, which typically have monolithic software
installations which are difficult to upgrade and maintain, into dedicated
network segments, isolated from the main corporate IT network. Based by this
recommendation, the U.S. Department of Veterans' Affairs created the
Medical Device Isolation Architecture
(MDIA)~\cite{VA/MedicalDeviceSecurity/2017}, which mandates the use of
software-based mechanisms to isolate medical devices and restrict their traffic
from entering the hospital's network. 

In this work, we analyze the effectiveness of logical network separation against
an attacker who has succeeded in installing a malicious agent on at least one of
the two separated network segments. The goal of the attacker is to communicate
with this agent, bypassing the security boundary enforced by a router or by
another network component.

\begin{figure}[t] 
  \begin{center} 
    \includegraphics[scale=0.27]{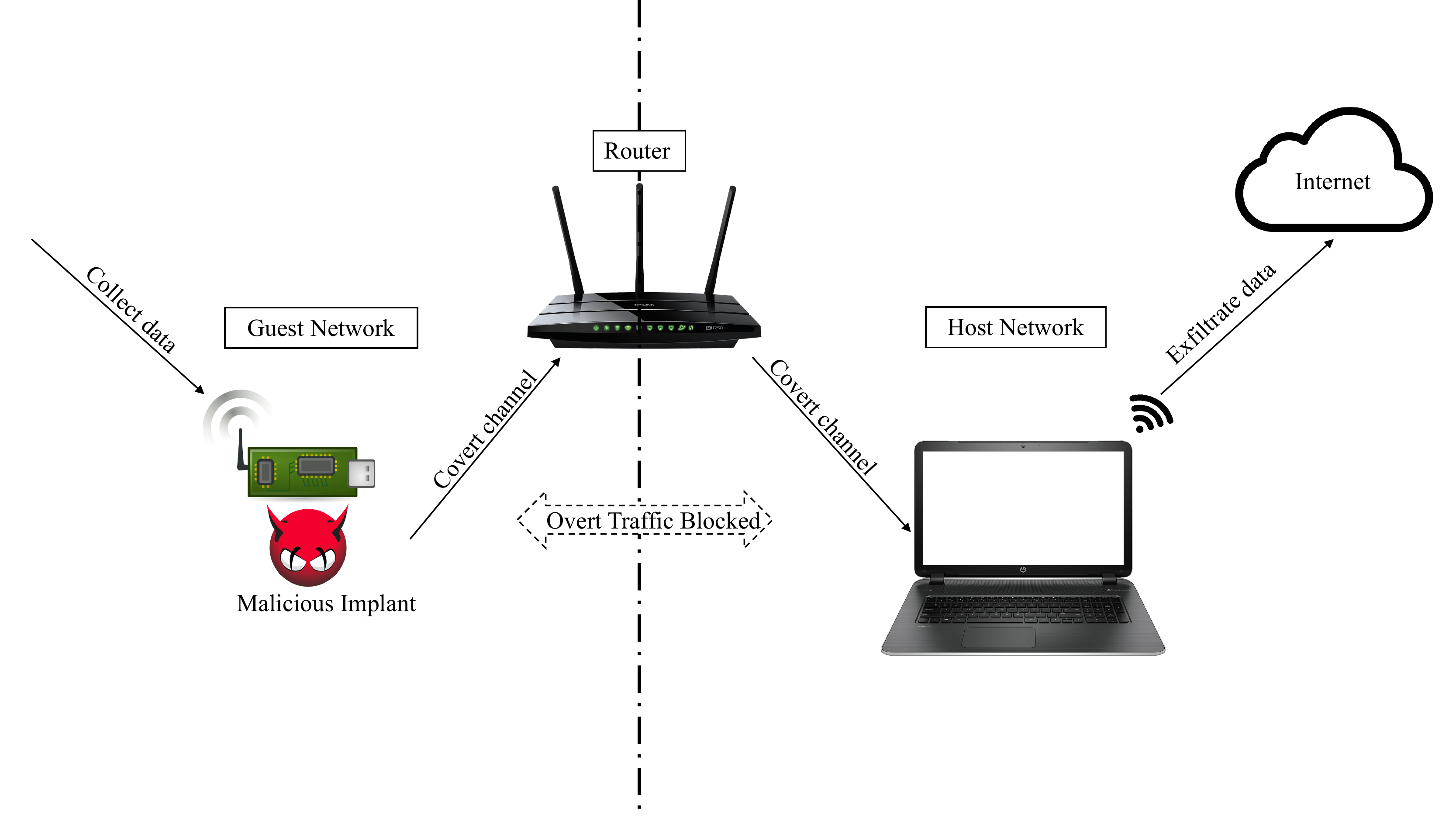} 
  \end{center}
  \caption{A covert channel between a host network and a guest network. Overt traffic is blocked, but the covert channel is not blocked. \label{f:covert-channel-host-guest}}
  \end{figure}

There are two important use cases in which a malicious actor may desire to
overcome this isolation: exfiltration and control. The exfiltration scenario is
illustrated in Figure~\ref{f:covert-channel-host-guest}. As shown in the Figure,
a malicious implant installed on the guest network has collected some sensitive
data, for example, a personal health-related sensor reading, and would like to
leak this data to the Internet. Only the host network, however, and not the
guest network, is connected to the Internet, and overt communications between
the two networks is blocked. As shown in the Figure, the malicious implant can
use a cross-router covert channel to send the data to the host network, and from
there to the Internet. In the control scenario, a remote command and control
server located on the Internet is interested in sending an activation command to
an advanced persistent threat (APT) installed on a device residing on the
sensitive host network. Only the guest network, however, and not the host
network, is connected to the Internet in this scenario.  The attacker can use a
cross-router covert channel to cause a computer residing on the guest network to
send the activation command to the implant. As we show in this article, our
attack can succeed even if the attacker has very limited permissions on the
infected device, and even an iframe hosting malicious JavaScript code can be
used for this purpose.



The general mechanism which is used for sending and receiving data in such a
restricted situation is called a \emph{covert channel}. As described in more
detail in Section~\ref{subsec:covert-channels}, a covert channel is a
communications link set up between two parties, the sender and the receiver, who
want to share some data between them where direct communication is not allowed.
In our particular case, we would like to exploit the fact that the router is a
shared resource between the host and the guest network, and use this router as
the covert channel. 

In this paper we make the following contributions:
\begin{itemize}
  \item We characterize cross-router covert channels, which allow leaking data
  between the host network and the guest network through the use of the router
  as a shared medium. We provide several metrics for the effectiveness of such
  channels, based on their pervasiveness, rate and covertness.
  \item We perform a survey of routers representing multiple vendors and
  price points, and identify a series of cross-router covert channels which
  impact some or all of the routers we survey.
  \item We classify the covert channels according to data rate, ease of
  identification and impact on network traffic.
  \item We discuss possible ways of identifying and preventing these
  cross-router leakages.
\end{itemize}

Understanding the limitations of software-based network isolation is very
important due to two factors: the first is the explosive growth in inexpensive
and relatively insecure IoT devices, and the second is the increasing dependence
of many organizations on a secure IT infrastructure.

\subsection{Covert Channels\label{subsec:covert-channels}}
Covert channels, first defined in 1973 by Lampson in
\cite{DBLP:journals/cacm/Lampson73}, are communication channels which exist
between two parties, a sender and a receiver, and can be used when overt
communication between these parties is prohibited due to privilege separation,
sandboxing or other architectural boundaries. In
\cite{DBLP:journals/comsur/ZanderAB07}, Zander et al. define two main two types
of covert channels: direct and indirect. A \emph{direct covert channel}
describes the case when the two parties run an innocuous-looking overt
communication channel, containing a hidden covert channel. An \emph{indirect
covert channel} describes the case when such an overt communication channel
between the parties does not exist. In this case, the two parties establish a
covert channel using some hardware which is shared between them.

In our particular case, a direct covert channel would correspond to a method of
direct data exchange between the host network and the guest network which is not
blocked by the router's isolation architecture. Direct covert channels can be
considered software bugs, and are relatively simple to fix in software, either
by the addition of additional firewall rules which block these data packets, or
by scrubbing the sensitive data and replacing it with random data. An indirect
covert channel, on the other hand, would be achieved by having the sender
selectively exhaust the finite hardware resources (CPU, memory, network
bandwidth, etc.) available on the router, and having the receiver measure the
effect of this varying resource consumption on its own performance. Blocking
this form of data transfer is more difficult, since it may require architectural
changes to the router.

We note that throughout this paper we do not consider communications channels
that are simply based on writing to a third-party server accessible to both the
host and the guest networks  (i.e. some shared resource on the Internet), since
this form of data exchange is relatively easy to detect and block. Furthermore,
in some network topologies either the host or the guest network cannot access
the Internet at all.

\subsection{Motivation} 
\begin{figure}[hbtp]
  \begin{center}
    \includegraphics[scale=0.39]{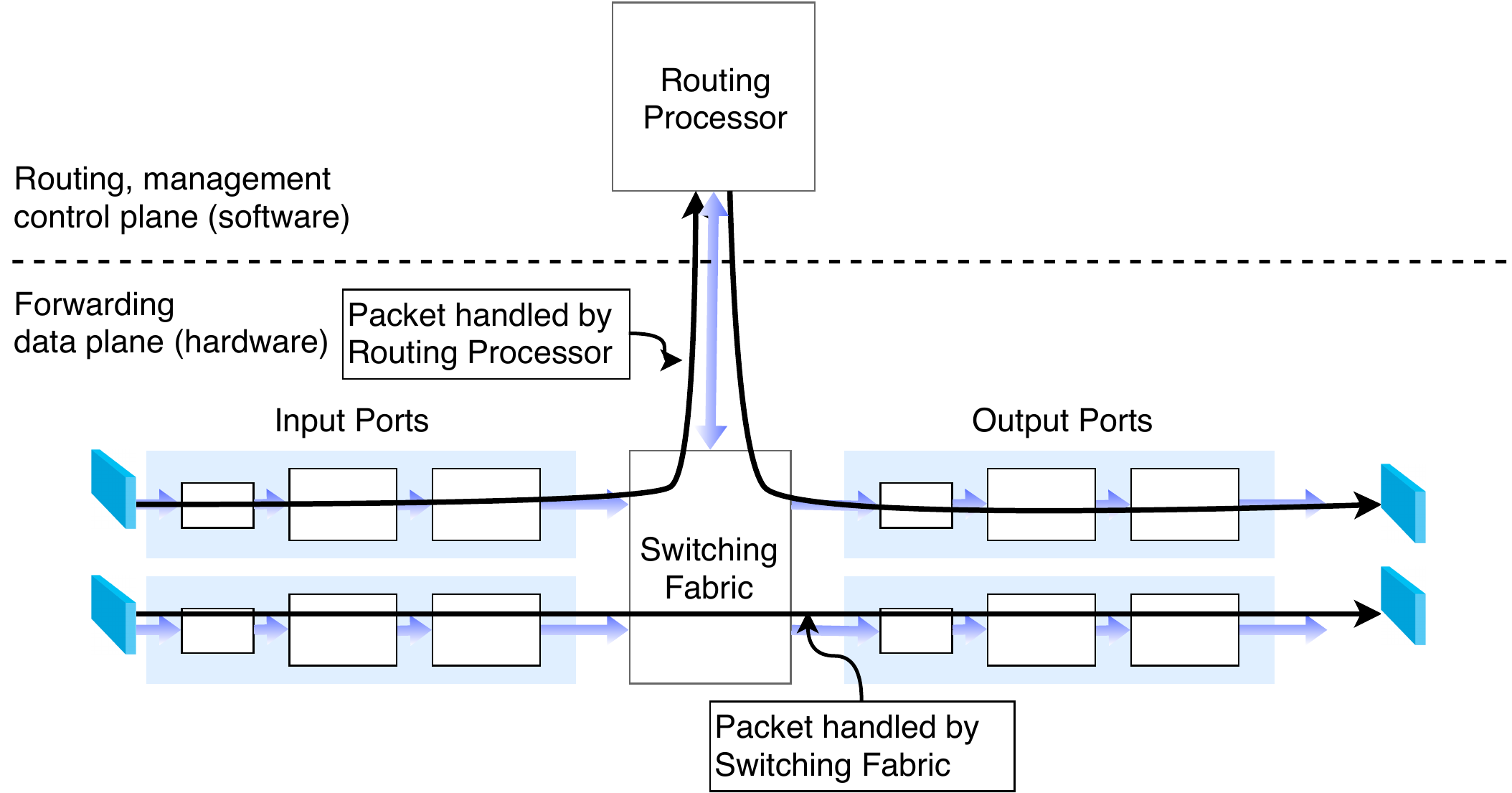}
  \end{center}
  \caption{Architecture of a router. Some packets are handled quickly by the switching fabric, while others are handled more slowly by the routing processor.}\label{f:router-architecture}
\end{figure}

The general architecture of a router, as described by Kurose and Ross
in~\cite{Kurose-Ross}, can be found in Figure~\ref{f:router-architecture}. As
noted in the Figure, the two main elements in the router are the software-based
control plane and the hardware-based routing plane. As Kurose and Ross note,
"the router’s input ports, output ports, and switching fabric together implement
the forwarding function and are almost always implemented in hardware. These
forwarding functions are sometimes collectively referred to as the \emph{router
forwarding plane}. \dots While the forwarding plane operates at the nanosecond
time scale, a router’s control functions —-- executing the routing protocols,
responding to attached links that go up or down, and performing management
functions  --— operate at the millisecond or second timescale. These
\emph{router control plane} functions are usually implemented in software and
execute on the routing processor (typically a traditional CPU)."

Guided by this discussion, we chose to focus our attempts to create a
cross-router covert channel not on the forwarding plane, which operates at line
speed, but rather on the slower control plane. We did so by generating traffic
which the router does not simply forward, but rather has to respond to in
software. While devices on the host network may have a wide variety of ways to
interact with the router's control plane, we claim that even the most
locked-down router must expose a bare minimum set of router control plane
functions to the guest network in order to function properly, notably the Dynamic Host Configuration
Protocol (DHCP)~\cite{RFC2131}, the Address Resolution Protocol
(ARP)~\cite{RFC0826}, and the Domain Name System (DNS)~\cite{RFC1034}. An
additional control plane feature which is often exposed to the guest network as
a convenience is the Internet Control Message Protocol~\cite{RFC0792}, commonly
used by the \texttt{ping} utility to verify network connectivity.

\section{Router-Based Covert Channels}
 
The covert channels we discovered can be broadly split into two groups: direct
covert channels and timing covert channels. In direct covert channels, the data
to be exchanged is directly encoded into a packet which is (erroneously) forwarded
between the host and guest networks. As soon as we discover such a form of
erroneously forwarded packet, using this form of covert channel is quite
straightforward. Timing-based covert channels, on the other hand, take advantage
of the shared resources on the router, such as CPU time, network and IPC
buffers, and so on. To exploit these channels, we need to construct
\textbf{sender and receiver gadgets} which cause an increased demand on the
router's control plane or sample this demand, respectively. Various combinations
of these sender and receiver gadgets can be used to form a covert channel,
depending on the router's support for different network protocols. In the
following Section we describe the direct covert channels we discovered, as well
as a series of sender and receiver gadgets used for timing-based covert
channels. In the next Section we show how these gadgets can be combined in
various ways to form complete covert channels.

\subsection{Direct Covert Channels}
\subsubsection{DHCP Direct}

\begin{figure}[hbtp]
  \begin{center}
    \includegraphics[scale=0.35]{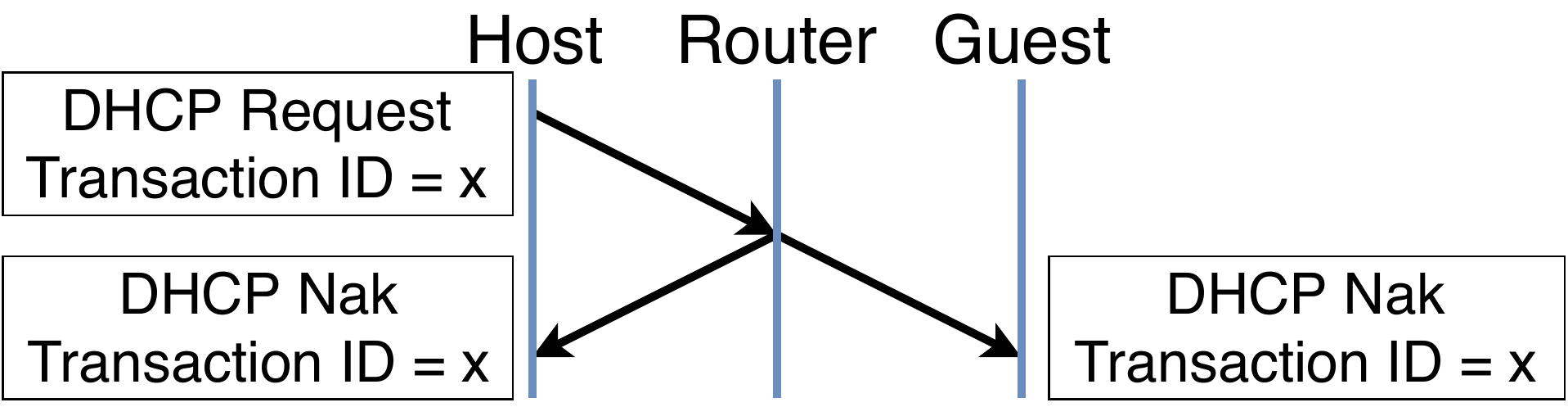}
  \end{center}
  \caption{The DHCP Direct covert channel. On some routers, a DHCP NAK from one network is erroneously sent to the other network.\label{f:dhck-nak-interaction-diagram}}
\end{figure}

The Dynamic Host Configuration Protocol (DHCP)~\cite{RFC2131} is a protocol used
to dynamically assign IP addresses and other network configuration parameters to
hosts joining a network. While the protocol formally involves a message exchange
between the host, or DHCP client, and a DHCP server present on the network, in
practice most residential and small business routers implement DHCP server
functionality themselves. The DHCP protocol begins with the client computer
sending a \textbf{DHCP Discover} message. The DHCP server will respond with a
\textbf{DHCP Offer} message, offering an IP address and other parameters. The
client then chooses an offer and sends a \textbf{DHCP Request} message with the
requested IP address and parameters. Finally, the DHCP server either sends a
\textbf{DHCP ACK} message affirming the requested IP and parameters, or a
\textbf{DHCP NAK} message denying the request. In today's reality of wireless
hosts joining and leaving the network in an ad-hoc manner, the DHCP server is
virtually mandatory in routers, and must be enabled both on the host and the
guest network.

The DHCP direct covert channel exploits the fact that some DHCP packets have an
unusual IP header, which includes 0.0.0.0 and 255.255.255.255 as the source and
destination addresses, respectively. This uncommon structure causes DHCP packets
to be handled by non-standard code paths on many devices. Furthermore, DHCP is
one of the protocols which must be supported on the guest network, since without
it network connectivity is impossible. On several of the routers we investigated,
the router responds to an invalid DHCP Request message sent from the guest
network with a broadcast DHCP NAK response which is sent to both the guest and
the host networks. Figure~\ref{f:dhck-nak-interaction-diagram} demonstrates how we
exploit this behavior to transfer data between these two networks. The arrows
between the participants in the attack describe the messages sent from one
participant to the other over time. The text by each arrow describes the message
sent between the participants. As we can see, a DHCP Request is sent to the
router with a certain Transaction ID field. Following the DHCP protocol, the
router responds with ACK/NAK message (in our case NAK), erroneously sending the
NAK to both Host and Guest networks with the same Transaction ID as found in the
DHCP Request. This allows encoding of data to be sent cross-router into the
32-bit Transaction ID field. 

\subsubsection{IGMP Direct}

\begin{figure}[hbtp]
  \begin{center}
    \includegraphics[scale=0.35]{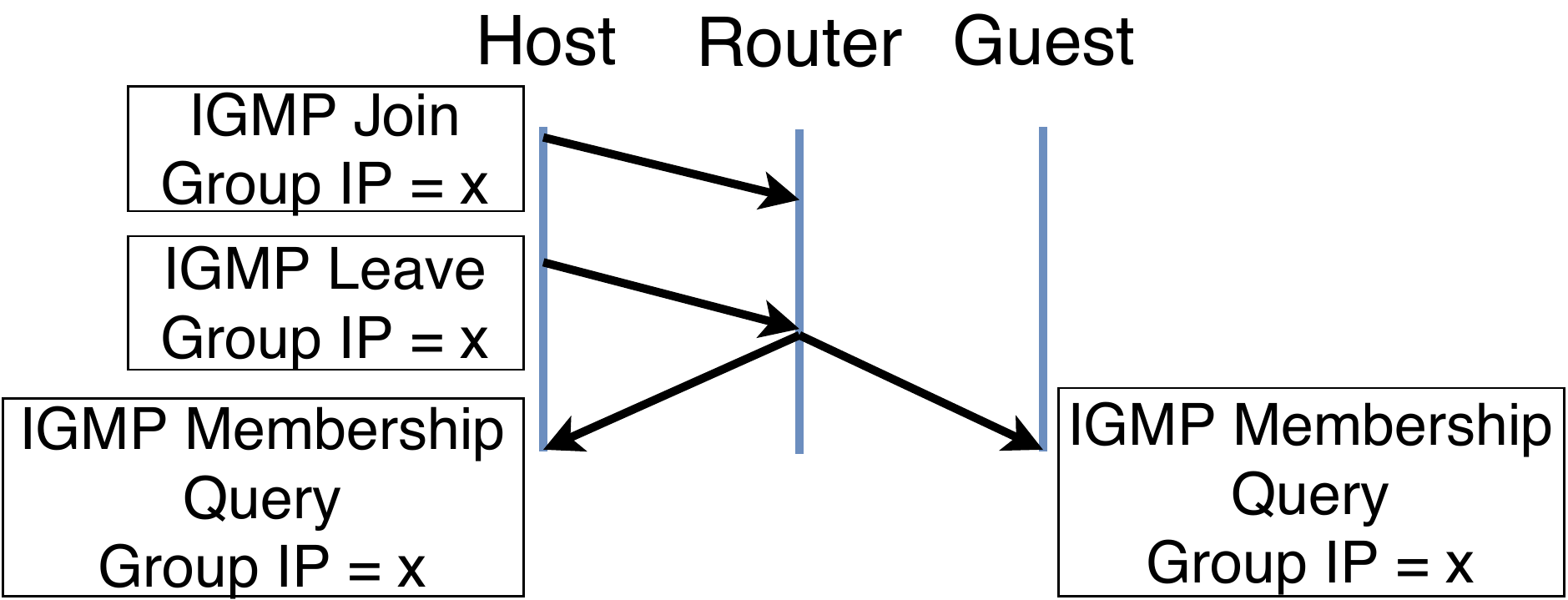}
  \end{center}
  \caption{The IGMP Direct covert channel. On some routers, an IGMP leave from one network erroneously causes an IGMP membership query to be sent to the other network.\label{f:igmp-leave-interaction-diagram}}
\end{figure}

The Internet Group Management Protocol (IGMP)~\cite{RFC2236} is a protocol used
on IPv4 networks to establish multicast group membership. Despite its extremely
limited use, this protocol is supported for historical reasons by a wide variety
of routers. According to the IGMP protocol, if a router discovers that the last
member of an IGMP group has left the group, it must check whether there are
remaining members in the group by sending an \textbf{IGMP Membership Query}
packet to all of its connected interfaces. 

The IGMP Direct covert channel exploits this property of the IGMP protocol. We
discovered that quickly joining and leaving a group from the host side caused an
IGMP Membership Query packet to be sent to both the host and guest networks on
routers TP1, TP2, DL2 and ED2. Figure \ref{f:igmp-leave-interaction-diagram} demonstrates
how we use this behavior to transfer data between these two networks. In
order to transfer data from the host network to the guest network, the sender
joins and then leaves an IGMP group. After it leaves, the router, following the
IGMP protocol, creates an IGMP Membership Query packet with the Group IP and
sends it to both the Host and the Guest networks. The data is transferred within
the Group IP field, which is completely controlled by the sender.
\subsubsection{ARP Direct} 
The Address Resolution Protocol (ARP)~\cite{RFC0826} is a link-level protocol
used to resolve a MAC address associated with an IP address. To resolve a MAC
address of a specific associated IP address, a station broadcasts an  
\textbf{ARP request} packet (a.k.a. "who-has") asking
for the MAC address of a station with a specific IP address in the
network. The station which has the IP address specified in the ARP request then
sends an \textbf{ARP response} packet (a.k.a. "is-at") with its own MAC address as an
answer.

ARP must always be enabled, even on the guest network, since it is used to locate the
router itself. We noticed, however, that some of the routers we evaluated
forwarded ARP requests, which are sent as broadcast packets, between the host
and the guest networks. Some routers restricted ARP forwarding only to requests
destined for the network's subnet mask, while some routers did not restrict this
traffic in any way. To use this leakage as a direct covert channel, the sender
can trivially issue an ARP request to an arbitrary computer on the network, using either
the lower 8 bits of the IP address, or the entire 32 bits in other cases, as the
data payload. 
\subsection{Timing Covert Channel Building Blocks}
The gadgets described in the following subsection can be used to either cause an
increased demand on the router's shared resources or to sample this demand. A
complete covert channel is formed by combining two of these gadgets, one on the
host network and one on the guest network.
\subsubsection{SSH}

\begin{figure}[hbtp]
  \begin{center} 
    \includegraphics[scale=0.35]{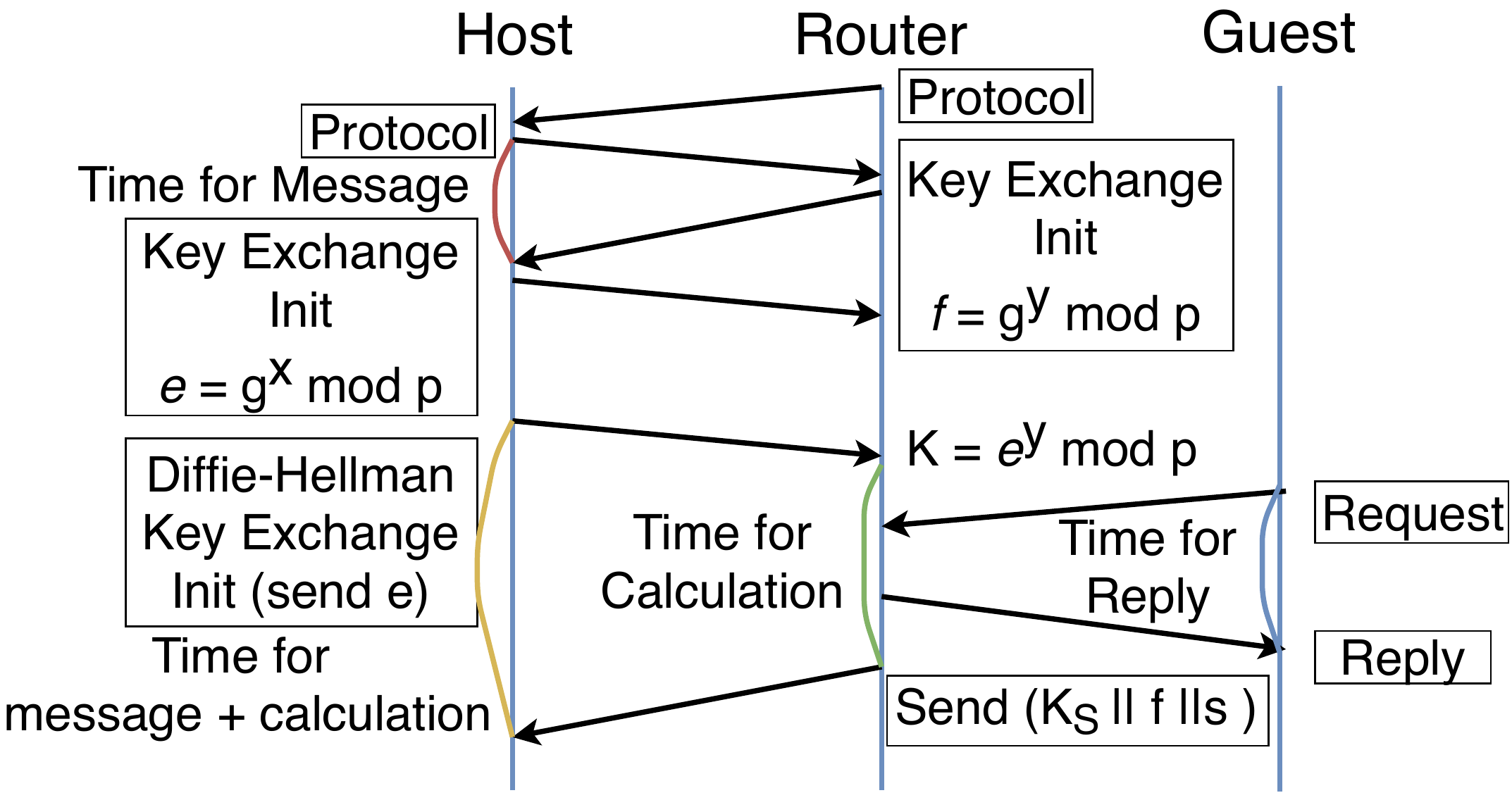}
  \end{center}
  \caption{The SSH Timing building block. Causing the router to perform an SSH Key Exchange makes it noticeably slower in responding to other requests.\label{f:ssh-kex-diagram}}
\end{figure} 

The SSH protocol~\cite{RFC4253} is used for remote access to various types of
network equipment, including several of the routers we evaluated in this work. We
take advantage of the fact that SSH connection setup is a relatively
CPU-intensive operation, specifically involving a modular exponentiation as part
of the key exchange process. Figure \ref{f:ssh-kex-diagram} describes the course
of the attack, with the three vertical lines in the interaction diagram
representing the Host , Router and Guest actors. As
shown in the Figure, the sender initiates SSH key exchange on the Host network.
The router carries out a modular
exponentiation as part of the key exchange,
and the connection is finally aborted before the protocol concludes. Since the
aborted connection stops before the authentication phase, we found that there is no evidence
in the router's log file that the connection attempt even occurred, adding to
the covertness of this attack.

To make sure the channel had a reasonable bit rate, we minimized the calculation
time by choosing the parameter set \texttt{diffie-hellman-group1-sha1}, which
has a small key size. This parameter set is, in fact, unsupported in
most modern implementations of SSH on desktops and servers, but can be enabled
using legacy mode command line parameters.

\subsubsection{CSRF}
Cross-Site Request Forgery, or CSRF, is a type of web attack described in RFC
6749~\cite{RFC6749} as "an exploit in which an attacker
causes the user-agent of a victim end-user to follow a malicious URI
(e.g., provided to the user-agent as a misleading link, image, or
redirection) to a trusting server (usually established via the
presence of a valid session cookie).". CSRF attacks have historically been used
to maliciously modify router settings, perform malicious bank transactions, and
so on. To prevent these attacks, modern browsers prevent cross-site read and write access
to websites unless a Cross-Origin Resource Sharing field is used. It is still,
however, possible for a website to display content from another website in an
embedded iframe\cite{RFC6454}.

To use CSRF as a timing covert channel, we take advantage of the fact that most
routers expose a web management interface on the host network. The attacker can
then coerce a victim on the host network into viewing an attacker-controlled web
page (for example, by showing a malicious advertisement), and access this
management interface using an embedded iframe element. Due to modern
CSRF protections built into most routers and browsers, it is rather difficult
for an adversary to maliciously change settings on the router using this method.
The timing channel, however, is still present, since web resources
requested by this iframe are served by the router's control plane. Therefore,
the CSRF channel can be used as a send gadget, by repeatedly loading the
router's webpage in an iframe, causing an increased CPU load on the router. It
can also be used as a receive gadget which gauges the load on the router by
measuring the time it takes for the router's management website to render (or,
in practice, to return an "access denied" error). 

\subsubsection{DHCP Timing}
The DHCP protocol, described above as a potential direct covert
channel, can also be used as a timing-based covert channel, even if it is
properly implemented. To use DHCP in this way, the attacker can send a valid
DHCP Request packet to the router and measure the time it takes for the router
to respond with a DHCP Acknowledge packet. This behavior is allowed by the DHCP
protocol, and is used for clients wishing to extend their leased address.

We noted experimentally that DHCP protocol interactions result in entries being
created on the log files of some routers. On one hand, this additional file
system activity increases the processing time of every DHCP transaction, making
it easier to use in a timing covert channel. On the other hand, this activity
leaves evidence which makes the attack easier to detect after the fact.

\subsubsection{ARP Timing}

As mentioned previously, the ARP protocol is enabled even on the guest network,
since it is used to locate the router itself. To use the ARP protocol as a
timing-based covert channel gadget, the sender repeatedly queries the router for
its own MAC address by sending an ARP who-has packet with the router's IP
address. The router has to answer this request, even on the guest network, for
the network to function. On the receiver side, the attacker sends an ARP request
again, and measures the time it takes for the router to respond to the ARP
request. As shown in the following Section, even a low rate of ARP requests, as
little as 100 packets per second, can affect the CPU load of the router in a
measurable way. 

\subsubsection{ICMP}

The Internet Control Message Protocol (ICMP)~\cite{RFC0792} is a supporting
protocol which is a vital part of the Internet Protocol suite. It is used to
provide feedback about problems and operational information in the networked
communication environment. One very common use for the ICMP protocol is the
\texttt{ping} command, which is used to diagnose network connectivity. When the
\texttt{ping} command runs, it sends an \textbf{ICMP echo request} packet to the
remote host, which is then expected to reply with an \textbf{ICMP echo reply}
packet. While support for ICMP is not mandatory, the ability to "ping the
router" is a common enough request for ICMP to be enabled on the guest networks
of some of the routers we evaluated.

As in the case of ARP, this protocol can be used as a timing-based covert
channel gadget by repeatedly sending ICMP echo requests to the router and then
measuring the time it takes for the router to respond.

\section{Methodology} 
To demonstrate the wide impact of the covert channel we discovered, we attempted
to reproduce our results on as many router models as possible, from multiple
vendors and price points. To prepare each router for experimentation, we first
inspected its online documentation, both on the official vendor website and on
the OpenWRT website, which contains hardware information for many router models.
Next, we made sure the router was factory reset and updated it to the most
recent firmware version we could find on the vendor's website. Then, we used the
router's web-based management interface to enabled the router's host and guest
isolation feature and connected two different computers to the router's host and
guest networks, respectively. We checked that the isolation feature works in
principle by verifying that na\"ive direct connections between the two computers
are blocked by the router.

Next, we attempted to identify any open services of the router by running the
\texttt{nmap} utility, both on the host and on the guest network. We also
passively monitored the router's activity by running \texttt{tcpdump} and
observing which services are advertised by the router, again both on the host
and on the guest network. Finally, we tested for the existence of direct or
indirect covert channels, by observing how the router reacted to a set of
uniquely-crafted messages which we found were more likely to breach the network
isolation feature, as described in the following sections.


\begin{figure}[hbtp]
  \begin{center}
    \includegraphics[scale=0.35]{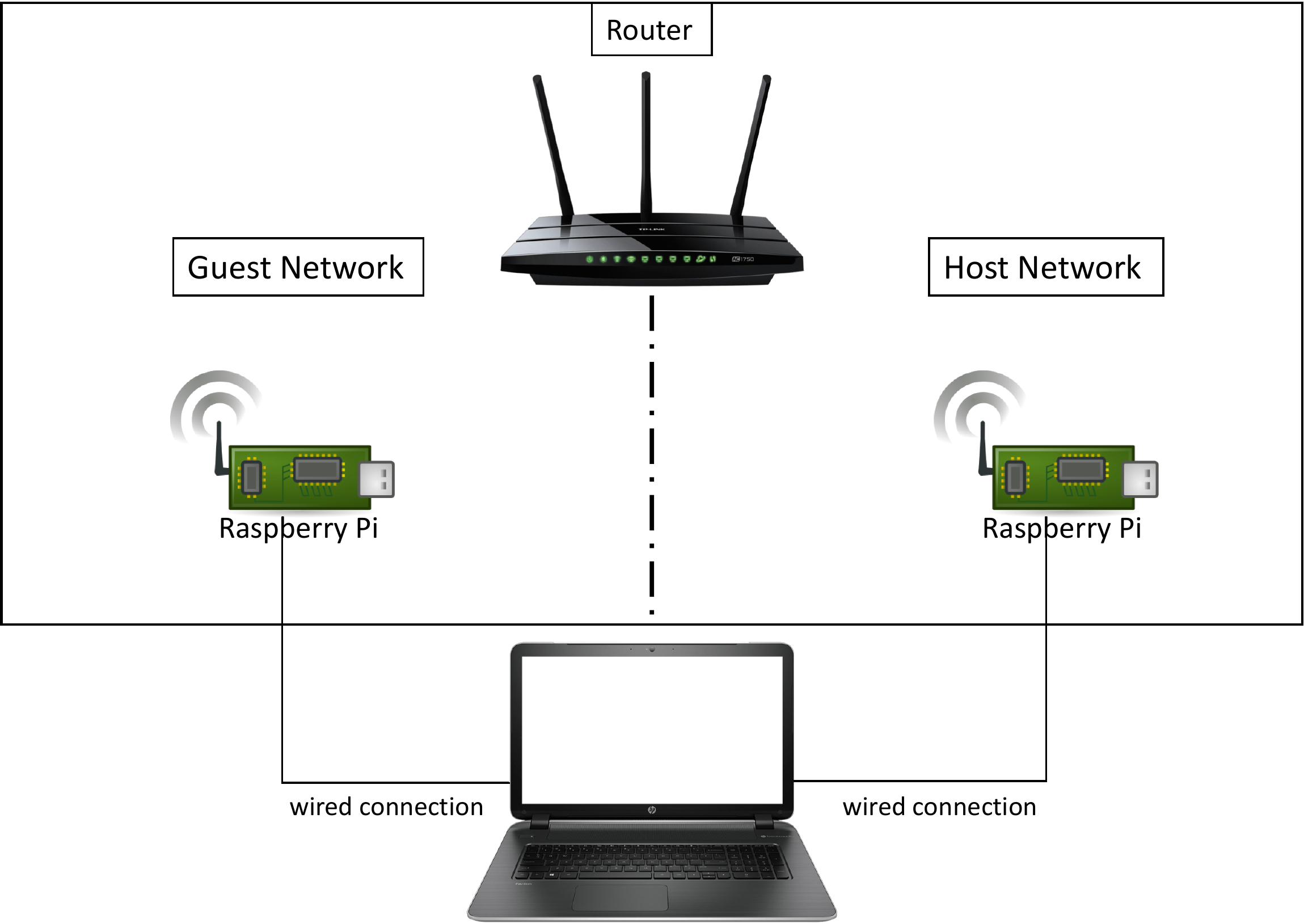}
  \end{center}
  \caption{Experiment Setup\label{f:experiment-setup}}
\end{figure}

Figure~\ref{f:experiment-setup} shows the experiment setup we used. As shown in
the Figure, each router under test was connected via wireless link to two
Raspberry Pi devices, one connected to the guest network and the other to the
host network. The two Raspberry Pis were in turn connected via a wired
Ethernet link to a test harness computer, which was used to start the
measurements and collect the experiment results. To make sure cross-router
communication was not achieved via an external third-party server, the router's
WAN/Internet port was left disconnected. 

\begin{table*}[t] {
  \centering
  \caption{Evaluated Routers\label{t:Evaluated-Routers}}
  \begin{tabular}{l|c|c|c|c|c|c|c}
    \toprule
    \textbf{Identifier} &\textbf{Vendor} & \textbf{Model} & \textbf{CPU type} & \textbf{Core count}
      & \textbf{CPU speed} & \textbf{Year introduced}& \textbf{Price}  \\
    \midrule
    TP1 & TP-Link           & Archer C3200 & Broadcom BCM4709A0 & 2 & 1 GHz &
    2015 & \$218   \\
    TP2 & TP-Link         &   Archer C2 & MediaTek MT7620A & 1 & 580 MHz   &2017
    & \$63 \\
    DL1 & D-Link           & DIR-882  & MediaTek MT7621A & 1 & 880 MHz & 2017 &
    \$154 \\
    DL2 & D-Link         & DIR-825AC   & Realtek RTL8197DN & 1 & 660 MHz   &
    2015 & \$50   \\
    ED1 & Edimax          & RG21S    &  MediaTek MT7621AT & 2 & 880 MHz & 2017 &
    \$209 \\
    ED2 & Edimax          & BR-6208AC   & Realtek RTL8881AQ & 1 & 520 MHz & 2014
    & \$47 \\
    LS1 & Linksys         & EA7500-eu  & Qualcomm IPQ8064 & 2 & 1.4 GHz & 2016 &
    \$185 \\
    \bottomrule
    \end{tabular}
  }
\end{table*}  

The list of routers we evaluated is listed in Table~\ref{t:Evaluated-Routers}.
As shown in the Table, the routers cover a variety of vendors and price points,
and have a wide diversity of CPU types, speeds and core counts.

\subsection{Criteria for channel quality}\label{s:critera-for-quality}
This report identifies many different types of covert channels. We propose the
following metrics to compare and evaluate the quality of each of the channels we
found.

The first and most significant criterion is the \textbf{pervasiveness} of the
channel: how widespread is this channel among the various types of hardware, and
how difficult would it be to fix this channel using a simple software upgrade. The
next criterion is the channel's \textbf{rate}: how much data can be transferred
per unit of time over this channel with a reasonable data rate. Finally, we can consider the
channel's degree of \textbf{covertness}: how similar is traffic sent using this
channel to regular traffic exchanged by the router, and how hard is this channel
to detect using forensic tools which examine log files and other external
artifacts. 
\section{Results}
\begin{table*}[htb] { 
  \begin{center}
  \caption{Covert Channels Supported by Different Routers}\label{t:covert-channel-support}
  \small The table summarizes the covert channels we found on each one of the
  routers described in Table~\ref{t:Evaluated-Routers}.

  The arrow direction describes the possible flow of the data
  between the guest (G) and host (H) networks.
  \begin{tabular}{lllllllll}
  \toprule \textbf{Channel} & \textbf{Type} & \textbf{TP1}& \textbf{TP2} & \textbf{DL1} & \textbf{DL2} & \textbf{ED1} & \textbf{ED2} & \textbf{LS1} \\
    \midrule 
    ARP-SSH   & Timing & G $\Rightarrow$ H & G $\Rightarrow$ H & -- & -- & -- & -- & --   \\
    ARP-ARP  & Timing & G $\Leftrightarrow$ H & G $\Leftrightarrow$ H & -- & G $\Leftrightarrow$ H & G $\Leftarrow$ H & G $\Leftrightarrow$ H & G $\Leftrightarrow$ H \\
    ARP-CSRF  & Timing & G $\Leftrightarrow$ H & G $\Rightarrow$ H & G $\Leftrightarrow$ H & G $\Rightarrow$ H & G $\Leftrightarrow$ H & G $\Rightarrow$ H & G $\Rightarrow$ H   \\
    ICMP-ICMP  & Timing & -- & -- & -- & -- & G $\Leftrightarrow$ H & -- & G $\Leftrightarrow$ H \\
    DHCP-ARP   & Timing & G $\Leftrightarrow$ H & G $\Leftrightarrow$ H & G $\Leftrightarrow$ H & G $\Leftrightarrow$ H & G $\Rightarrow$ H & G $\Leftarrow$ H & G $\Leftrightarrow$ H \\
    DHCP Direct  & Direct & G $\Leftarrow$ H  & G $\Leftarrow$ H & -- & G $\Leftrightarrow$ H & -- & G $\Leftrightarrow$ H & -- \\
    IGMP Direct  & Direct & G $\Leftarrow$ H & G $\Leftarrow$ H & -- & G $\Leftrightarrow$ H & -- & G $\Leftrightarrow$ H & --  \\
    ARP Direct  & Direct & G $\Leftarrow$ H & G $\Leftarrow$ H & -- & G $\Leftrightarrow$ H & -- & G $\Leftrightarrow$ H & --  \\
    \bottomrule \end{tabular}
  \end{center}
      }   
\end{table*}

Table~\ref{t:covert-channel-support} lists the types of attacks we evaluated on our routers, either by
immediately applying direct covert channels, or by combining send and receive
gadgets for timing-based covert channels. When discussing timing-based covert
channels, the first gadget is always used on the guest side, and the second
gadget on the host side. We indicate that a timing-based covert channel exists only if
Student's Independent two-sample $t$-test, applied to the outputs from the
receive gadget, can tell apart between two sets of 1,000 timings, obtained either with and
without the sender gadget, with a significance of $p<0.05$. 

The direction of the arrows indicates whether we discovered a host-to-guest covert channel, a guest-to-host covert
channel, or a bidirectional channel spanning both directions.

In addition to the channels examined in this report, we note that
timing-based channels exist for additional combinations of sender and receiver
gadgets, for example DHCP vs. DHCP, using similar mechanisms
to the ones described above. Discussion of these channels is omitted for space.

\subsection{DHCP Direct}

\begin{figure}[btp]
  \begin{center}
    \includegraphics[scale=0.6]{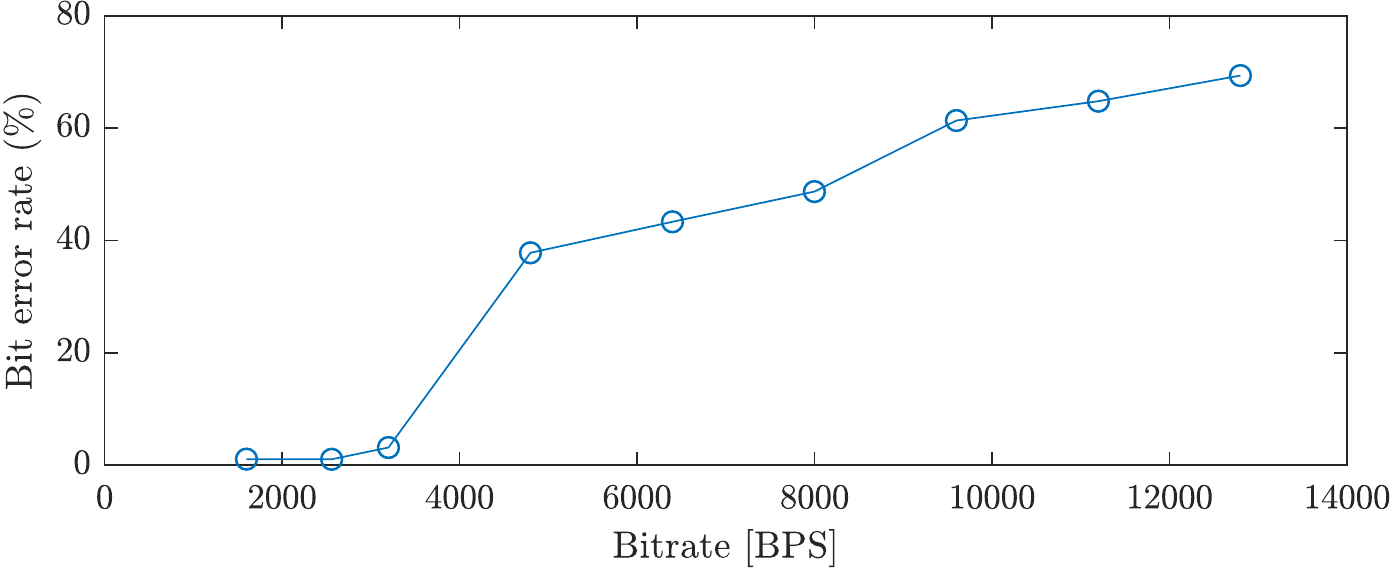}
  \end{center}
  \caption{DHCP Direct error rate by bit rate (TP2)\label{f:dhcp-error-bit-rate}}
\end{figure}

In order to measure the performance of the DHCP attack, Figure
\ref{f:dhcp-error-bit-rate} describes the bit error rate as a function of the
sending bit rate. The circles are the actual
output and the line emphasizes the change in the error rate over the bit rate.
We can see that up to around 3200 bits per second the bit error rate remains low
and above that the router is flooded and cannot handle all the DHCP Requests,
 leading to a jump in the bit error rate above 3200 bits per second.

We implemented a simple end-to-end attack demonstration in Python which provides
a chat functionality between two isolated networks using this covert channel. It
is relatively straightforward to extend this demonstration to a full
bidirectional pipe which can carry higher-level traffic such as PPP or SSH
tunnels~\cite{maurice2017hello}.
\subsection{IGMP Direct}

\begin{figure}[hbtp]
  \begin{center}
    \includegraphics[scale=0.6]{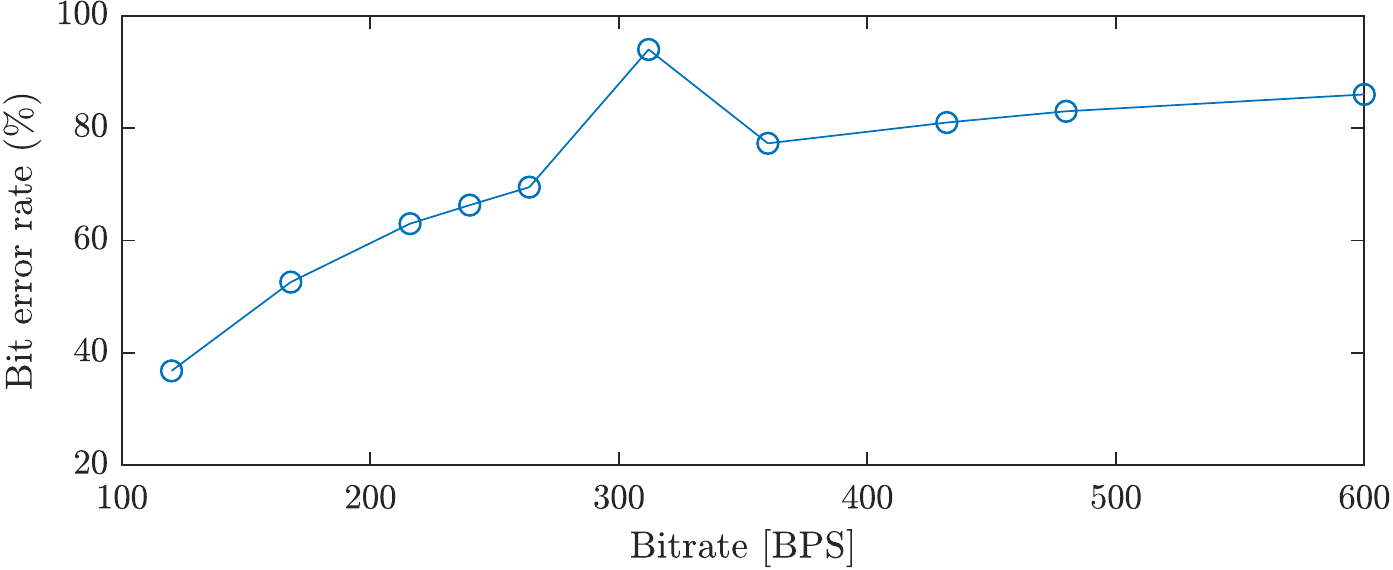} 
  \end{center}
  \caption{IGMP Direct error rate by bit rate (TP2)\label{f:igmp-error-bit-rate}}

\end{figure}

Figure~\ref{f:igmp-error-bit-rate} describes the bit error rate as function of
the bit rate of the IGMP attack described in
\ref{f:igmp-leave-interaction-diagram}.  We can see that the error rate grows with the bit
rate, and that above 170 bits per second the error rate becomes
than 50 percent.
 
\subsection{ARP vs. SSH}

\begin{figure}[hbtp]
  \begin{center}
    \includegraphics[scale=0.45]{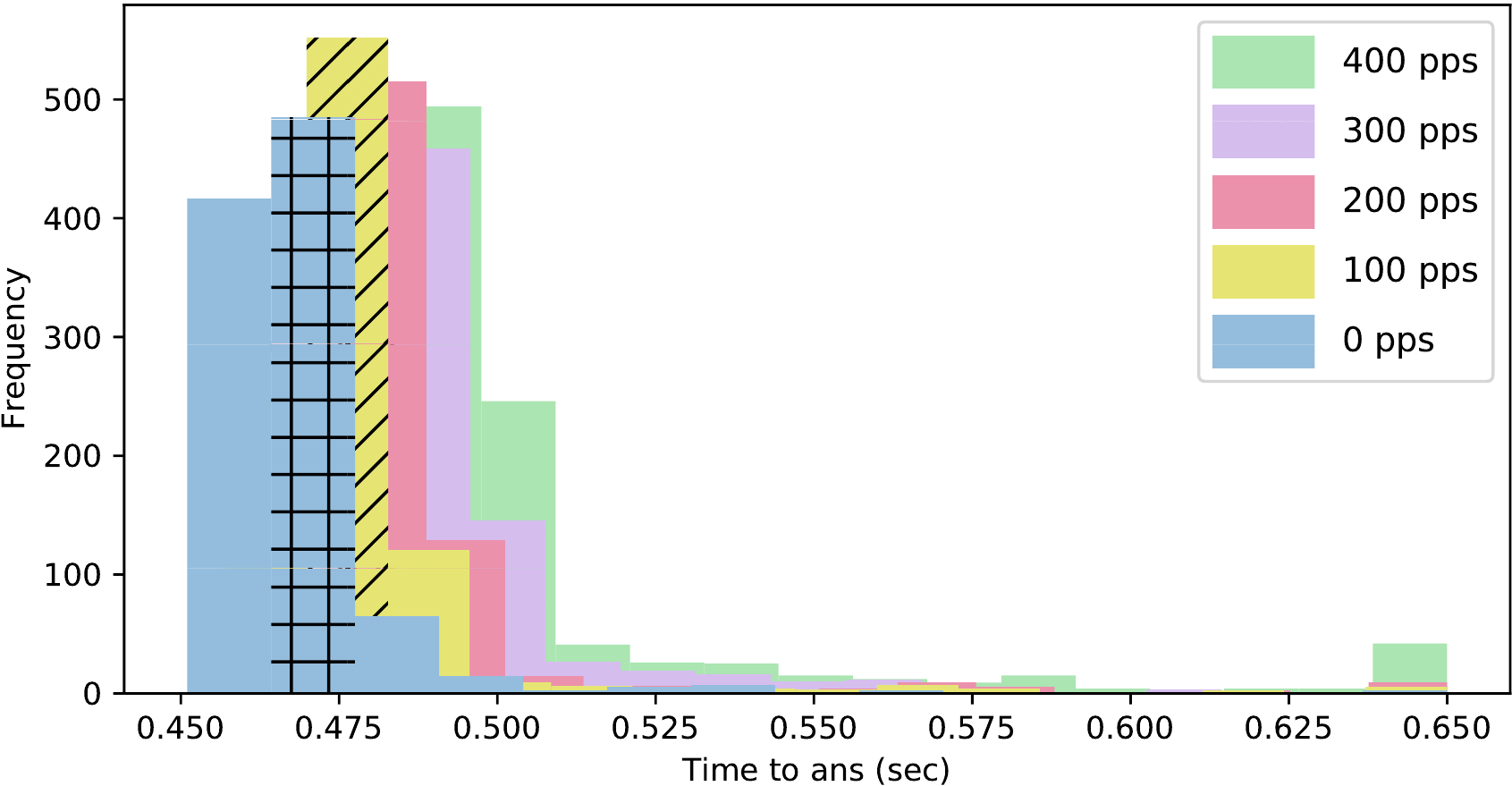}
  \end{center}
  \caption{ARP vs. SSH timing attack (TP2).\label{f:ssh-vs-arp-timing}
  Each histogram describes a different test-run in which the receiver sends
  1,000 SSH requests, while the sender sends a stream of ARP requests in different rates,
  between 0 to 400 packets per second.}
\end{figure}

To demonstrate the performance of the SSH vs. ARP attack we present Figure
\ref{f:ssh-vs-arp-timing}, which was captured on router TP2.
The figure compares the time it took the router to answer
the receiver's SSH requests while the sender sent ARP requests and while it
did not. Each color describes a different test-run in which the receiver sends
1,000 SSH requests, while the sender sends a stream of ARP requests in different rates,
between 0 to 400 packets per second. The blue bar is the test-run in which the
sender sends no ARP requests. The other bars are the test-runs in which the
sender sends ARP requests in different rates, from 100 to 400 packets per
second. It can be seen that there is a clear difference between the
measurements, and that the router takes more time to answer the receiver when
the sender is sending ARP requests.

\subsection{ARP vs. ARP}

\begin{figure}[btp]
  \begin{center}
    \includegraphics[scale=0.45]{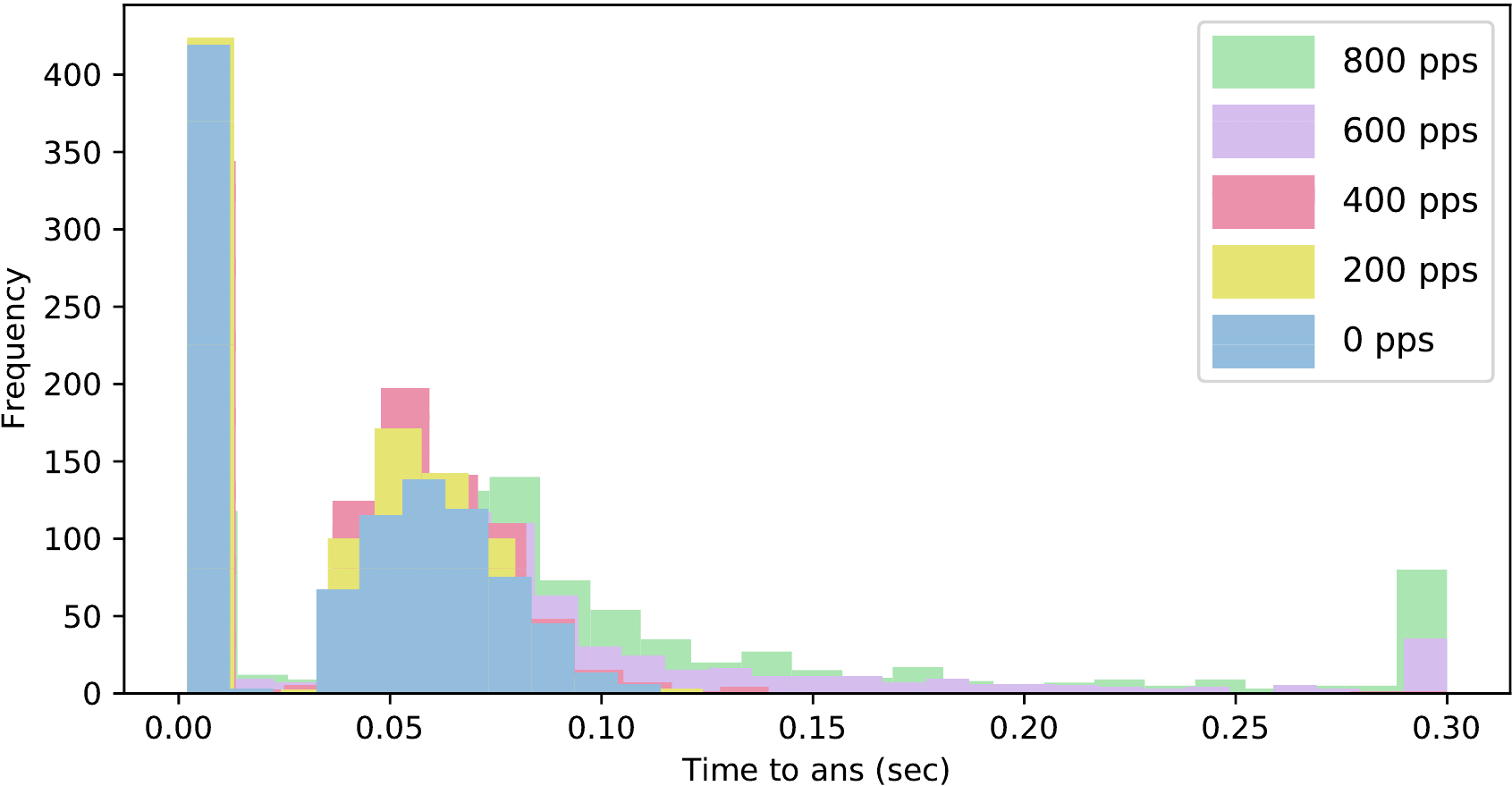} 
    \\[1\baselineskip]
    \includegraphics[scale=0.45]{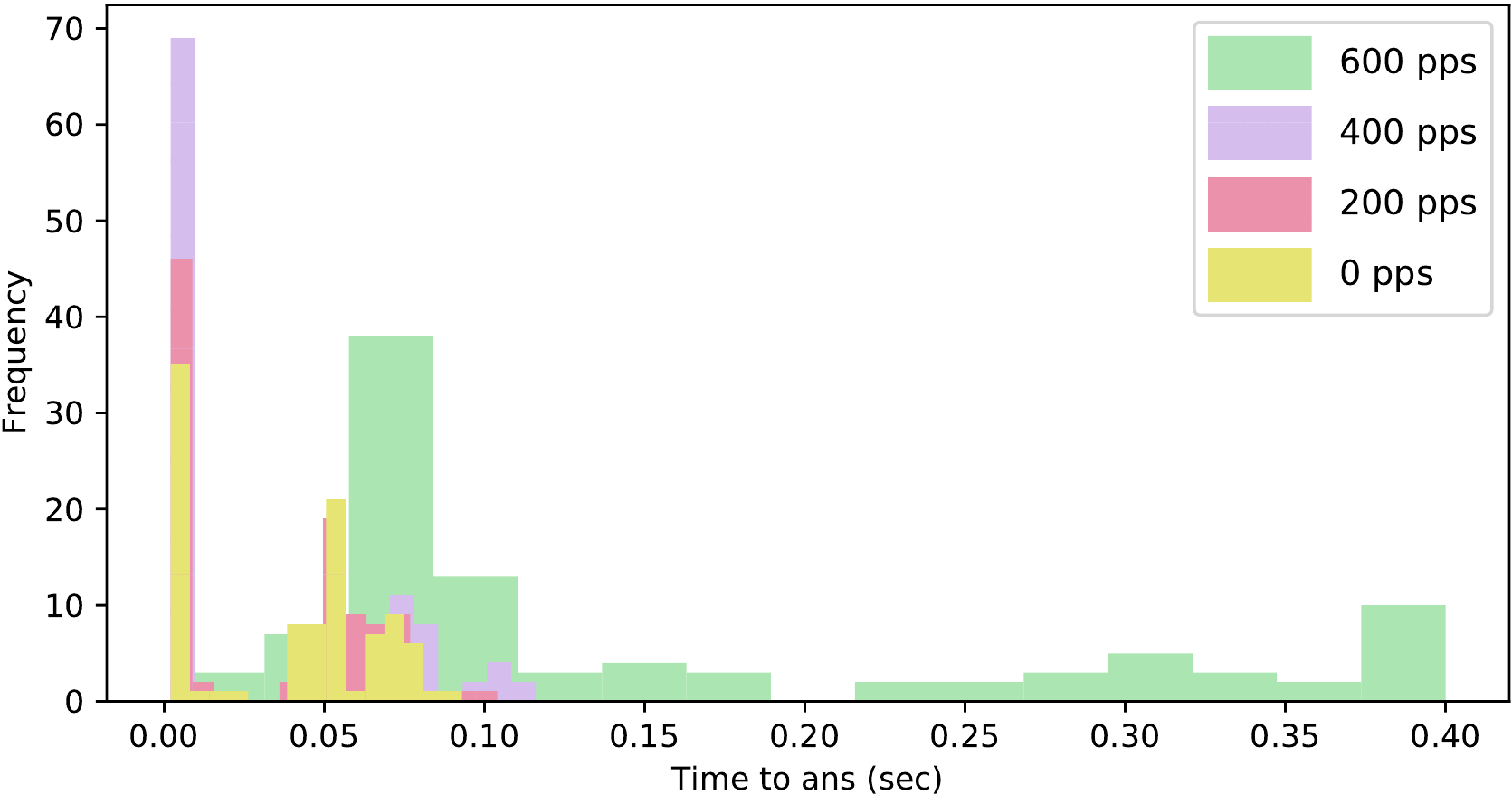}
  \end{center}
  \caption{ARP vs ARP timing attack (top: ED2, bottom: TP2)\label{f:arp-vs-arp-timing}
  Each histogram describes a different test-run in which the receiver sends 1000 ARP
  requests and the sender continuously sends ARP requests in different rates, from
  0 to 800 packets per second.}
\end{figure}

Figure \ref{f:arp-vs-arp-timing} shows measurements of the time it took the
router to answer the receiver's ARP requests, sent from the host network, while
the sender sent ARP requests, sent from the guest network, and while it did not.
The following Figure shows this attack on two different routers: ED1 and TP2.
Each color describes a different test-run in which the receiver sends 1000 ARP
requests and the sender continuously sends ARP requests in different rates, from
0 to 800 packets per second. On the bottom graph it is clearly seen that when
the sender sends ARP requests at a high rate, it takes more time for the router
to answer the receiver. On the other hand, the top graph illustrating router ED1
shows no significant difference in the response time, but the t-test performed
on the results shows significant difference between the different test-runs (p-value
 lower than 0.05).

\subsection{ICMP vs. ICMP}

\begin{figure}[hbtp]
  \begin{center}
    \includegraphics[scale=0.45]{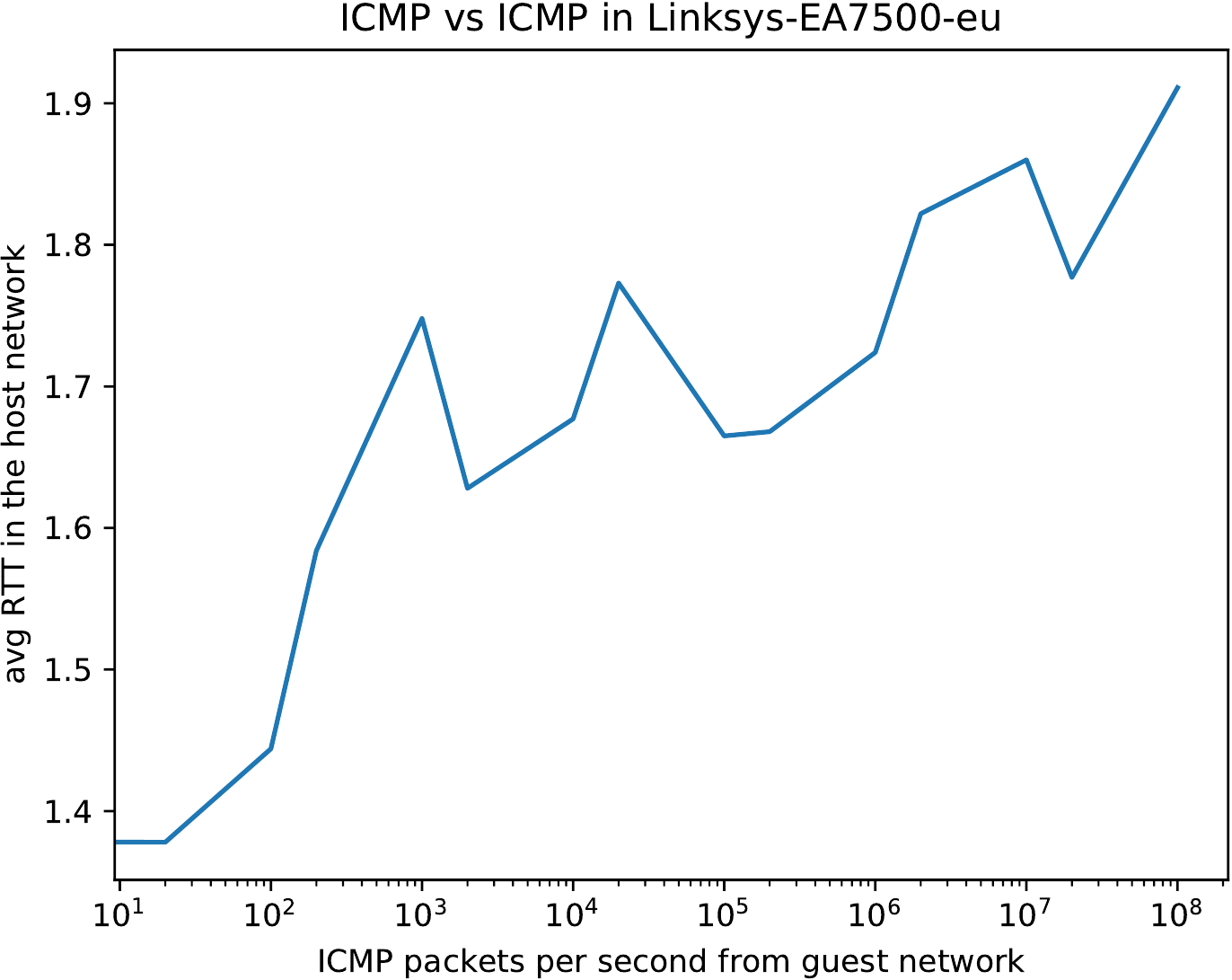} 
  \end{center}
  \caption{Average round-trip time for an ICMP request on the host network, as
  function on
  the rate of ICMP packets sent from the guest network
  (LS1).\label{f:ICMP-ICMP-timing}}
\end{figure} 

Figure \ref{f:ICMP-ICMP-timing} shows the average round-trip time for an ICMP
request measured at the host network, as a function of the rate of ICMP
packets sent from the guest network, as measured on router LS1.  We can see that the measured
response time on the host network goes up with the number of packets per second
sent on the guest network, allowing data to be sent between these two networks.

\subsection{ARP vs. CSRF}

\begin{figure}[hbtp]
  \begin{center} 
    \includegraphics[scale=0.45]{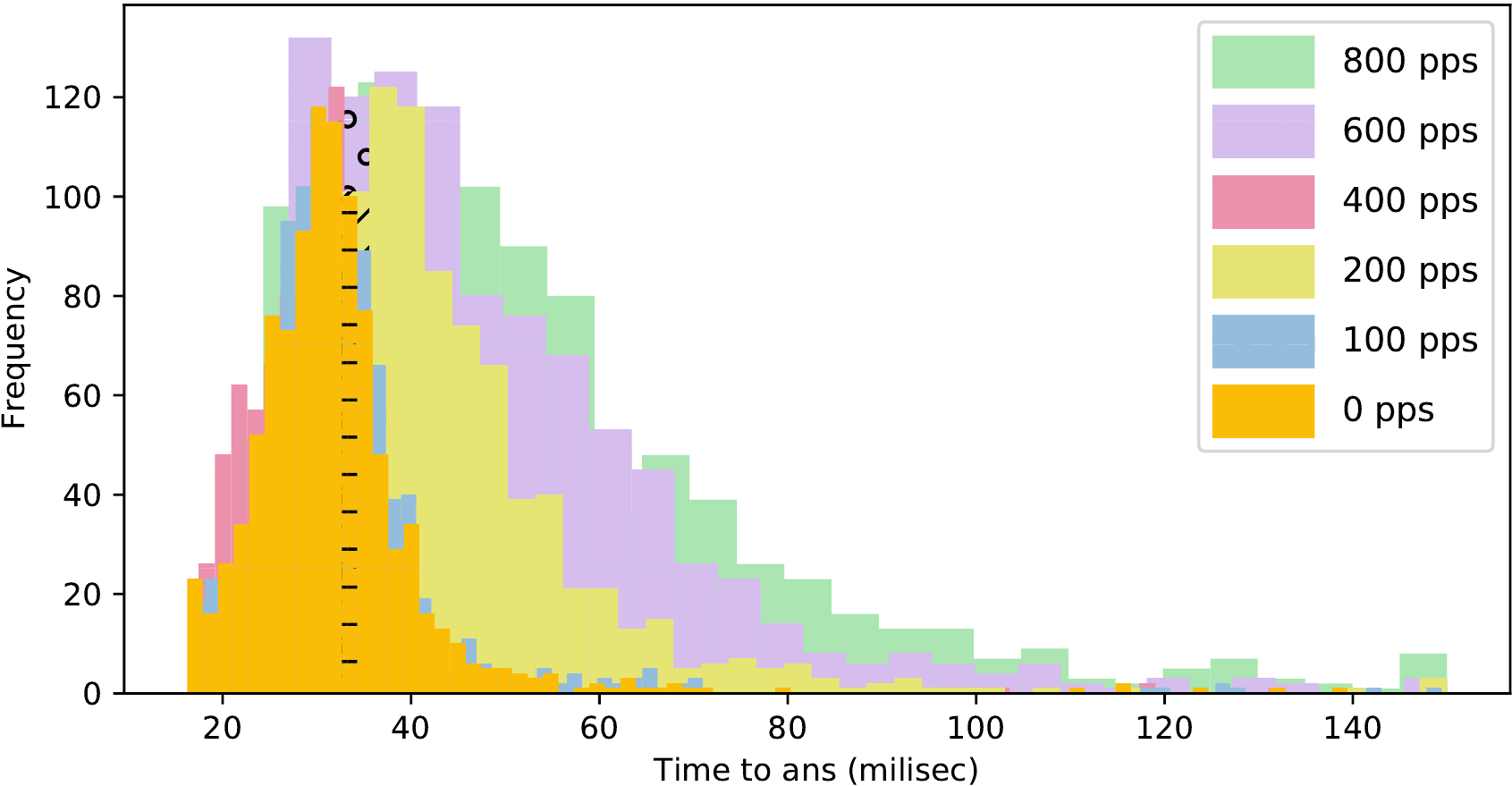}
  \end{center}
  \caption{ARP vs. CSRF timing attack (DL1)\label{f:ARP-CSRF-timing}
  Each histogram shows the round-trip time for an CSRF
  request measured using JavaScript on the host network, as a function of the rate of ARP
  packets sent from the guest network.}
\end{figure} 

Figure \ref{f:ARP-CSRF-timing} shows the round-trip time for an CSRF
request measured using JavaScript on the host network, as a function of the rate of ARP
packets sent from the guest network, as measured on router DL1.  As shown in the
Figure, a timing-based covert channel can be established in this case even
without using any custom software or hardware on the host side. Therefore this
attack can work even if the node in the protected network is not compromised.

\section{Discussion}

\begin{table}[htb] {
  \centering
  \caption{Quality of Different Covert Channels}\label{t:results-summary}
  \begin{tabular}{lllll}
    \toprule
      \textbf{Channel}  & \textbf{Pervasiveness} 
                                 & \textbf{Rate} & \textbf{Covertness}  \\
    \midrule
    ARP-SSH              & ++    & ++   & +   \\
    ARP-ARP              & +++   & +    & +++ \\
    ARP-CSRF             & +++   & +    & ++  \\
    ICMP-ICMP            & ++    & ++   & ++  \\
    DHCP-ARP             & +++   & ++   & + \\
    DHCP Direct          & +     & +++  & ++  \\
    IGMP Direct          & +     & +++  & +   \\
    ARP Direct           &  ++   & +++  & ++  \\
    \bottomrule
    \end{tabular}
      }
\end{table}

Table~\ref{t:results-summary} summarizes the quality of each of the covert
channels we identified, according to the criteria proposed in
Subsection~\ref{s:critera-for-quality}. Of all the channels we evaluated, the
two direct channels, IGMP direct and DHCP direct, have the highest data rate and
can be used to transfer thousands of bits per second. On the other hand, their
existence is due to a bug in the router implementations and they are therefore
easy to fix. In addition, IGMP traffic is extremely rare in production networks,
and DHCP activity generates log file entries, making both channels limited in
their covertness. The SSH-ARP channel is one of the most covert of the timing-based
channels we identified, as it generates no log-file entries since the SSH
connection establishment never concludes. We still consider it less pervasive
than the other timing-based covert channels due to the limited amount of routers
with default support for the SSH protocol. The two ARP-based channels, CSRF-ARP
and ARP-ARP, are the most pervasive in our opinion, since virtually all routers
expose some sort of web server on their host network side, and all routers
support the ARP protocol. The CSRF-ARP channel is slightly less stealthy since
the thousands of web requests per second may constitute an irregular access
pattern which can be detected by external intrusion detection systems. Both
ARP-based channels are limited in their rate because ARP packets are easily
handled by the router's CPU and generate only a minimal resource footprint.
Finally, the ICMP-ICMP channel is both more covert and more stealthy than the
ARP-ARP channel, but its pervasiveness is limited by the fact that not all
routers expose ICMP on the guest network side.


\subsection{Related Work}
 
In 1973 Lampson first introduced covert channels in the context of monolithic
systems, as a mechanism by which a process at a high security level, leaks
information to a process at a low security level, where the low-security process
would not otherwise have access to this information
\cite{DBLP:journals/cacm/Lampson73}. Various types of covert channels are
presented below.

\subsubsection{Timing covert channels}
Network timing channels transfer information using packet arrival patterns and
 not by the actual content of the packet. Network timing channels can be divided
 into two primary types: Timing Channel and Sorting Channel. The former uses
 predefined timing intervals to obtain information, reception of a packet
 represents a ’1’ and the absence of a packet indicates a ’0’
 \cite{DBLP:conf/ccs/CabukBS04}, while the latter uses the order in which the
 packets has arrived to build the message transferred \cite{ahsan2002covert}. In
 the timing channel implemented by Cabuk et al., the covert information is
 being divided into small fixed-size parts. The sender and the receiver
 synchronize by using a \emph {Special Start Sequence} in the beginning of every
 frame. They also suggest a detection method against these kind of covert
 channels which was found to be highly efficient in detecting such covert
 channels even under random noise or changing time intervals. The timing channel
 they presented uses a direct channel to convey information. Maurice et al. have
 successfully established an indirect covert channel between virtual machines
 running on different cores \cite{DBLP:conf/dimva/MauriceNHF15} . They exploit
 the inclusive feature of caches, which is a shared resource among the machines,
 allowing a core to evict lines in the private first level cache of another
 core. By measuring access times to the cache, the receiver notices which bit
 was received. In our study we perform a similar manipulation on the CPU of a
 router.

\subsubsection{Network covert channels}

Several papers presented the ability to carry storage covert channel by
modifying the headers of different protocols of layers of the Internet Protocol
stack. For example, Rowland~\cite{DBLP:journals/firstmonday/Rowland97} presented
a covert channel based on modifying the IP Identification Field, the TCP Initial
Sequence Number Field and the TCP Acknowledge Sequence Number Field. In another
example, Handel~\cite{handel1996hiding} presented a series of covert channels
based on the OSI network model. Many variations on these kinds of covert
channels exist, based on different sections of the network packet headers.
These channels are also very easy to eliminate using methods that scrub or
remove these headers~\cite{DBLP:conf/uss/HandleyPK01}.
In~\cite{DBLP:journals/csur/WendzelZFH15} Wendzel et al. provide a survey of hiding methods
in network covert channels. The methods we
investigated in this paper are unique in the fact that they make use of the
shared medium of the router to construct the covert channel, an idea which is
novel to the best of our knowledge. 

As for timing covert channels carried over the network, the data is transferred
by creating delays in packet transmission, dropping packets or reorganizing
them. Handel et al.~\cite{handel1996hiding} introduced a covert channel that
drops or delays packets by jamming the medium of in the CSMA/CD protocol using
bit per packet modulation to send the data. In another example, Ogen et
al.~\cite{ogen2018sensorless} presented a covert channel over the 802.11
protocol, using the Clear Channel Assessment in the 802.11 protocol to delay
packets transmission within few milliseconds and measuring that small delay
using JavaScript at the application layer. In contrast to the channel developed
by ~\cite{ogen2018sensorless}, the channel described in this paper does not require the sender
to by physically close to the router. In addition, the covert channel described
here does not require any custom sender hardware, and can be implemented in
software only on consumer products. One advantage of
~\cite{ogen2018sensorless} over the channel described here is that their covert
channel does not require knowledge of the Wi-Fi network's password.

\subsection{Detection and Prevention}

There are two general approaches for defending against covert channels:
detecting activity on a potential channel, and interfering with the covert
channel to the point of completely blocking it. Both approaches require
awareness of potential covert channels and often imply some degradation in
performance. Limitation and auditing countermeasures have also been discussed in
several other works.

Detecting a timing or storage covert channel requires a profile of the channel's
activity in a ``benign'' setting, i.e. without an attack in progress, and
measurements of its activity to test for data transfer on the channel. The basic
idea is that in timing and storage channels, data transfer is achieved by
contention on a shared resource. Both types of channels require nodes from both
network segments to repeatedly poll the resource, often at higher rate than the
background rate for such polling. For example, in our experiments the background
rate for ARP requests was lower than one packet per second, while the rate
during an attack\footnote{Note that we tried to maximize throughput in these
attacks.} was greater than $1000$ packets per second and with a high incidence
of concurrent activity from the same nodes in separate segments. Both rate and
concurrent activity can often be measured and correlated, raising an alarm.
General anomaly detection algorithms can be used to detect activity on a direct
side-channel, with effectiveness that is governed by the background activity of
this channel.

The detection approach suffers from an inherent disadvantage: false positives or
false negatives in detection. The reason is that users of the covert channel can
reduce the rate of the channel and thereby reduce the difference between the
activity on the channel when it is used and when it is not. 

Methods for prevention of a covert channel depend on the type of the channel.
Direct covert channels should be viewed as system bugs and should be corrected
by hardware or software vendor or by better configuration of network devices.
Storage side-channels are very difficult to block and should be avoided as much
as possible as part of system design.

A comprehensive approach to blocking timing side-channels is to divide the
router's computing resources into time slices, statically allotting time slices
to each network segment. Requests from nodes in a certain network segment will
be served only during time slices allotted to that segment. The advantage of
this approach is that timing side-channels are almost completely blocked, since
activity in one network segment does not affect activity in other segments. The
disadvantage of the static time slot method is that performance decreases as the
router is less flexible in serving requests. A different approach to interfere
with, although not completely prevent, timing side-channels uses the channels'
sensitivity to the distribution of time measurements for received messages.
Therefore, a router that adds a random delay before sending a message will
effectively increase the error probability in decoding and therefore decrease
the rate of the channel. Note that the router is constrained in the magnitude of
the delays it adds, i.e. in the channel error it introduces, since it needs to
serve legitimate customers in reasonable time. 

\subsection{Responsible Disclosure}
We sent a draft of our findings to the manufacturers of the routers listed in
Table~\ref{t:Evaluated-Routers} during May 2019. During June 2019 the Belkin/Linksys
security response team notified us that they do not intend to fix the
vulnerability we disclosed. None of the other router vendors responded to our
disclosure. Our vulnerability reports for the various channels and models were
granted CVE IDs CVE-2019-13263, CVE-2019-13264, CVE-2019-13265, CVE-2019-13266,
CVE-2019-13267, CVE-2019-13268, CVE-2019-13269, CVE-2019-13270 and
CVE-2019-13271.

\subsection{Conclusion}

In this work we showed that logical network isolation based on host and guest
networks can be overcome by the use of specially-crafted network traffic. All of
the routers we surveyed are vulnerable to at least one class of cross-router
covert channel, and fixing this vulnerability is far from trivial. A
hardware-based solution seems to be the only way of guaranteeing isolation
between secure and non-secure network devices.

\section*{Acknowledgments}
This research was supported by Israel Science Foundation grants 702/16 and
703/16. The authors would like to thank Cl\'{e}mentine Maurice for inspiring
this research, and our shepherd Paul Pearce for helping us improve the paper.

\bibliographystyle{plain}
\bibliography{Cross-Router-Covert-Channel.bib}

\begin{thebibliography}{10}

\bibitem{VA/MedicalDeviceSecurity/2017}
Medical device security, {VA} enterprise design patterns privacy and security,
  January 2017.

\bibitem{ahsan2002covert}
Kamran Ahsan.
\newblock Covert channel analysis and data hiding in tcp/ip.
\newblock {\em Canada, University of Toronto}, 2002.

\bibitem{RFC6454}
A.~Barth.
\newblock The web origin concept.
\newblock RFC 6454, RFC Editor, December 2011.
\newblock \url{http://www.rfc-editor.org/rfc/rfc6454.txt}.

\bibitem{DBLP:conf/ccs/CabukBS04}
Serdar Cabuk, Carla~E. Brodley, and Clay Shields.
\newblock {IP} covert timing channels: design and detection.
\newblock In Vijayalakshmi Atluri, Birgit Pfitzmann, and Patrick~D. McDaniel,
  editors, {\em Proceedings of the 11th {ACM} Conference on Computer and
  Communications Security, {CCS} 2004, Washington, DC, USA, October 25-29,
  2004}, pages 178--187. {ACM}, 2004.

\bibitem{RFC2131}
Ralph Droms.
\newblock Dynamic host configuration protocol.
\newblock RFC 2131, RFC Editor, March 1997.
\newblock \url{http://www.rfc-editor.org/rfc/rfc2131.txt}.

\bibitem{RFC2236}
William~C. Fenner.
\newblock Internet group management protocol, version 2.
\newblock RFC 2236, RFC Editor, November 1997.
\newblock \url{http://www.rfc-editor.org/rfc/rfc2236.txt}.

\bibitem{handel1996hiding}
Theodore~G Handel and Maxwell~T Sandford.
\newblock Hiding data in the {OSI} network model.
\newblock In {\em International Workshop on Information Hiding}, pages 23--38.
  Springer, 1996.

\bibitem{DBLP:conf/uss/HandleyPK01}
Mark Handley, Vern Paxson, and Christian Kreibich.
\newblock Network intrusion detection: Evasion, traffic normalization, and
  end-to-end protocol semantics.
\newblock In Dan~S. Wallach, editor, {\em 10th {USENIX} Security Symposium,
  August 13-17, 2001, Washington, D.C., {USA}}. {USENIX}, 2001.

\bibitem{RFC6749}
D.~Hardt.
\newblock The oauth 2.0 authorization framework.
\newblock RFC 6749, RFC Editor, October 2012.
\newblock \url{http://www.rfc-editor.org/rfc/rfc6749.txt}.

\bibitem{Kurose-Ross}
James Kurose and Keith Ross.
\newblock {\em Computer Networking: A Top-Down Approach (7th Edition)}.
\newblock Pearson, 2016.

\bibitem{DBLP:journals/cacm/Lampson73}
Butler~W. Lampson.
\newblock A note on the confinement problem.
\newblock {\em Commun. {ACM}}, 16(10):613--615, 1973.

\bibitem{DBLP:conf/dimva/MauriceNHF15}
Cl{\'{e}}mentine Maurice, Christoph Neumann, Olivier Heen, and Aur{\'{e}}lien
  Francillon.
\newblock {C5:} cross-cores cache covert channel.
\newblock In Magnus Almgren, Vincenzo Gulisano, and Federico Maggi, editors,
  {\em Detection of Intrusions and Malware, and Vulnerability Assessment - 12th
  International Conference, {DIMVA} 2015, Milan, Italy, July 9-10, 2015,
  Proceedings}, volume 9148 of {\em Lecture Notes in Computer Science}, pages
  46--64. Springer, 2015.

\bibitem{maurice2017hello}
Cl{\'e}mentine Maurice, Manuel Weber, Michael Schwarz, Lukas Giner, Daniel
  Gruss, Carlo~Alberto Boano, Stefan Mangard, and Kay R{\"o}mer.
\newblock Hello from the other side: Ssh over robust cache covert channels in
  the cloud.
\newblock {\em NDSS, San Diego, CA, US}, 2017.

\bibitem{RFC1034}
P.~Mockapetris.
\newblock Domain names - concepts and facilities.
\newblock STD~13, RFC Editor, November 1987.
\newblock \url{http://www.rfc-editor.org/rfc/rfc1034.txt}.

\bibitem{ogen2018sensorless}
Rom Ogen, Kfir Zvi, Omer Shwartz, and Yossi Oren.
\newblock Sensorless, permissionless information exfiltration with wi-fi
  micro-jamming.
\newblock In {\em 12th $\{$USENIX$\}$ Workshop on Offensive Technologies
  ($\{$WOOT$\}$ 18)}, 2018.

\bibitem{RFC0826}
David~C. Plummer.
\newblock Ethernet address resolution protocol: Or converting network protocol
  addresses to 48.bit ethernet address for transmission on ethernet hardware.
\newblock STD~37, RFC Editor, November 1982.
\newblock \url{http://www.rfc-editor.org/rfc/rfc826.txt}.

\bibitem{RFC0792}
J.~Postel.
\newblock Internet control message protocol.
\newblock STD~5, RFC Editor, September 1981.
\newblock \url{http://www.rfc-editor.org/rfc/rfc792.txt}.

\bibitem{DBLP:journals/firstmonday/Rowland97}
Craig~H. Rowland.
\newblock Covert channels in the {TCP/IP} protocol suite.
\newblock {\em First Monday}, 2(5), 1997.

\bibitem{DBLP:journals/cacm/SametingerRLO15}
Johannes Sametinger, Jerzy~W. Rozenblit, Roman~L. Lysecky, and Peter Ott.
\newblock Security challenges for medical devices.
\newblock {\em Commun. {ACM}}, 58(4):74--82, 2015.

\bibitem{stouffer2011guide}
Keith Stouffer, Joe Falco, and Karen Scarfone.
\newblock Guide to industrial control systems (ics) security.
\newblock {\em NIST special publication}, 800(82):16--16, 2011.

\bibitem{DBLP:journals/csur/WendzelZFH15}
Steffen Wendzel, Sebastian Zander, Bernhard Fechner, and Christian Herdin.
\newblock Pattern-based survey and categorization of network covert channel
  techniques.
\newblock {\em {ACM} Comput. Surv.}, 47(3):50:1--50:26, 2015.

\bibitem{RFC4253}
T.~Ylonen and C.~Lonvick.
\newblock The secure shell (ssh) transport layer protocol.
\newblock RFC 4253, RFC Editor, January 2006.
\newblock \url{http://www.rfc-editor.org/rfc/rfc4253.txt}.

\bibitem{DBLP:journals/comsur/ZanderAB07}
Sebastian Zander, Grenville~J. Armitage, and Philip Branch.
\newblock A survey of covert channels and countermeasures in computer network
  protocols.
\newblock {\em {IEEE} Communications Surveys and Tutorials}, 9(1-4):44--57,
  2007.

\bibitem{DBLP:conf/soca/ZhangCWHCS14}
Zhi{-}Kai Zhang, Michael Cheng~Yi Cho, Chia{-}Wei Wang, Chia{-}Wei Hsu,
  Chong~Kuan Chen, and Shiuhpyng Shieh.
\newblock Iot security: Ongoing challenges and research opportunities.
\newblock In {\em 7th {IEEE} International Conference on Service-Oriented
  Computing and Applications, {SOCA} 2014, Matsue, Japan, November 17-19,
  2014}, pages 230--234. {IEEE} Computer Society, 2014.

\end{thebibliography}

\end{document}